\documentclass[12pt]{iopart} 
\usepackage{amsfonts}
\usepackage{graphicx}
\expandafter\let\csname equation*\endcsname\relax 
\expandafter\let\csname endequation*\endcsname\relax 
\usepackage{amsmath}
\usepackage{amssymb}
\usepackage{bbm}
\usepackage{mdframed}
\usepackage{mathtools}
\usepackage{cite}
\usepackage{aas_macros}
\usepackage[colorlinks=true]{hyperref}
\newcommand{\mE}{\mathcal{E}}
\newcommand{\mB}{\mathcal{B}}

\begin{document}

\title{On the Ambiguity in Relativistic Tidal Deformability}

\address{\ \\ Samuel E. Gralla \\ Department of Physics \\ University of Arizona }

\begin{abstract}
The LIGO collaboration recently reported the first gravitational-wave constraints on the tidal deformability of neutron stars.  I discuss an inherent ambiguity in the notion of relativistic tidal deformability that, while too small to affect the present measurement, may become important in the future.  I propose a new way to understand the ambiguity and discuss future prospects for reliably linking observed gravitational waveforms to compact object microphysics. 
\end{abstract}

\maketitle 

\renewcommand*{\thefootnote}{\arabic{footnote}} 
\setcounter{footnote}{0}

\vspace{-5mm}
\section{Introduction}

Just two years after the beginning of operations, the advanced Laser Interferometer Gravitational-Wave Observatory (LIGO) is not only meeting, but vastly exceeding expectations.  Binary black hole (BH) mergers have been louder and more frequent than expected \cite{LIGOdiscovery,LIGO2,LIGO3} and the recent spectacular detection of a neutron star (NS) merger \cite{LIGONS}, complete with electromagnetic emission across the entire spectrum \cite{LIGONSMM}, has inaugurated a new era of multi-messanger astronomy.  Among the most exciting of upcoming possibilities is the prospect of studying the fundamental properties of matter at nuclear density by constraining the neutron star equation of state \cite{flanagan-hinderer2008}.  The equation of state determines the tidal deformability of neutron stars, which in turn influences the phasing of the gravitational waveform.  The first step towards realizing this dream was taken in Ref.~\cite{LIGONS}, where posterior distributions for certain tidal deformability parameters $\Lambda_1$ and $\Lambda_2$ were reported.

Ref.~\cite{LIGONS} also took the second step of comparing these results to theoretical calculations of the deformability $\Lambda$.  In this paper I will argue that the parameter $\Lambda$ of theoretical calculations is not necessarily the \textit{same} $\Lambda$ appearing in waveform models.  The origin is a fundamental ambiguity in the notion of relativistic tidal deformability.  The discrepancy is almost certainly much smaller than the uncertainty in this first measurement, but nevertheless demands attention as we move into the era of gravitational-wave astronomy.  As measurement accuracy improves, it will become necessary to confront the ambiguity before measurements of $\Lambda_1$ and $\Lambda_2$ can be translated into precise constraints on the behavior of matter at nuclear density. 

The basic problem of principle is well known \cite{damour5PN,damour-nagar2009,yagi-yunes2014,favata2014}: in the Post-Newtonian (PN) equations of motion for a compact binary, the tidal deformability parameters first appear at 5PN order, where the remaining terms are unknown.  A rigorous demonstration of the leading effects of tidal deformability will have to await further developments in the PN expansion.  In the meantime, one can pursue the simpler strategy of adding in 
the terms that can be derived for an \textit{intrinsic} quadrupole, replacing the intrinsic quadrupole moment with the induced quadrupole moment as characterized by $\Lambda_1$ and $\Lambda_2$ (e.g. \cite{flanagan-hinderer2008} and references therein).  This provides an observational definition of $\Lambda_1$ and $\Lambda_2$, which is effectively that recently used by LIGO. 

Unfortunately, it is not obvious how this observational definition connects back to theoretical definitions.  For a Newtonian star, there is no ambiguity in the theoretical definition and it seems clear that it will agree with the observational one.  A main purpose of this note is to call attention to an ambiguity in the case of a relativistic star that calls this identification into doubt.  Because there is no preferred coordinate system in a relativistic system, one cannot define the quadrupole moment as an integral over the mass density in the usual way.  For isolated stars one can instead define a quadrupole moment by the asymptotic behavior of the field, but this strategy fails for induced quadrupole moments, where the tidal field perturbation is not asymptotically flat.  
Previous work \cite{hinderer2008,binnington-poisson2009,damour-nagar2009,kol-smolkin2012} has fixed the ambiguity by adopting a preferred choice of coordinate system or a preferred solution of the linearized Einstein equation.

Ultimately, these ambiguities can only be cleared up by performing the whole calculation of the waveform, complete with neutron star microphysics, in a single computational framework.  In the mean time, I suggest here a simple way of understanding the ambiguities and thereby framing the discussion.  The key point is that while the perturbed metric of a compact object immersed in a weak tidal field is not asymptotically flat, the \textit{difference} between the metric perturbations of two different objects with the same mass is indeed asymptotically flat.  The multipole moments of this ``difference metric'' can then be invariantly defined in the usual way \cite{geroch1970,hansen1974,thorne1980,gursel1983}, and might be called the ``induced moment differences''.  This makes clear that the ambiguity is not so bad as it first might seem; it is just a single overall constant for each moment.

How then, should one fix the free constant?  The standard definition corresponds to the insistence that black holes do not develop induced quadrupoles.  Personally, I do not find this choice particularly compelling.  A spinning black hole develops a quadrupole moment like any other body, and one would expect a tidal field to have the same effect.  Furthermore, the vanishing of a coefficient with no associated symmetry is unnatural from the effective field theory point of view \cite{porto2016}.  However, there is no obvious alternative choice of coefficient, and ultimately, one's preferred definition is simply a matter of taste.

Of course, matters of taste do not influence the outcomes of experiments.  My position, therefore, is that one should regard the standard induced moments as encoding the \textit{difference} between the tidal response of a neutron star and that of a black hole of the same mass, and attempt to move forward using this understanding.  In particular, one could imagine considering the \textit{difference} between the full 5PN equations of motion for binary black holes and for binary neutron stars.  It is then plausible that, though the full expression is unknown, upon taking the difference only the tidal terms will survive, with only the induced quadrupole moment \textit{difference} appearing.  This could be attributed to the theoretically-defined induced quadrupole moment difference with some degree of confidence.  Going further, one could attempt to measure the value of the observationally-defined $\Lambda$ for a black hole and thereby fix the free constant.  In this way, neutron star waveforms are consistently calibrated with respect to black hole physics.  A similar strategy was pursued recently in Ref.~\cite{barkett-etal2016}.

The remainder of this paper is organized as follows.  In Sec.~\ref{sec:review}, I review the Newtonian approach and the relativistic ambiguity.  In Sec.~\ref{sec:relativistic}, I more carefully analyze the ambiguity in a relativistic framework and estimate its size.  In Sec.~\ref{sec:differences}, I discuss the notion of tidal deformability differences.  In Sec.~\ref{sec:comparison}, I compare to previous work.  In Sec.~\ref{sec:waveform}, I discuss effects on the waveform.

\section{Newtonian Definition and Relativistic Ambiguity}\label{sec:review}

In Newtonian physics, the induced quadrupole is just the quadrupole moment of the density perturbation.  Consider immersing an initially static, spherically symmetric body in an external tidal field,
\begin{align}\label{Phiext}
\Phi_{\rm ext}=\tfrac{1}{2} \mE_{ij} x^i x^j,
\end{align}
where $\mE_{ij}$ is symmetric and trace-free.  Let $\delta \rho$ denote the density perturbation developed by the body in response.  Then the induced quadrupole moment is simply
\begin{align}\label{QNewt}
Q_{ij} = \int d^3x \delta \rho\left( x_i x_j - \tfrac{1}{3} \delta_{ij} \right).
\end{align}
By definition, $Q_{ij}$ is also symmetric and trace-free. The tidal deformability $\lambda$ is then defined by
\begin{align}\label{lambda}
Q_{ij} = - \lambda \mE_{ij}.
\end{align}
This parameter $\lambda$ has dimensions of $(mass) (length)^2 / (time)^2$.  Since only the radius $R$ of the star and Newton's constant $G$ are involved in the calculation, the natural dimensionless combination is 
\begin{align}\label{k}
k = \tfrac{3}{2} G \lambda R^{-5},
\end{align}
where the factor of $3/2$ is conventional.  The constant $k$ is called the (quadrupolar) tidal Love number. 
The recent LIGO analysis \cite{LIGONS} instead scaled out the gravitational radius $GM/c^2$, defining
\begin{align}\label{Lambda}
\Lambda = \tfrac{2}{3} k \mathcal{C}^{-5}, \qquad \mathcal{C} \equiv \frac{G M}{c^2 R}.
\end{align}
Here $\mathcal{C}$ is the stellar compactness.  This parameter $\Lambda$ is natural in the relativistic setting since the stellar radius does not appear, i.e. $\Lambda=\lambda/M^5$ in units where $G=c=1$.  We henceforth set $G=c=1$.

For a relativistic star, the definition \eqref{Phiext}-\eqref{lambda} becomes ambiguous due to the coordinate freedom of general relativity: When integrating over the body in Eq.~\eqref{QNewt}, which coordinate system is to be used?  For isolated bodies, this problem can be solved by instead defining the moment according to the asymptotic behavior of the field \cite{geroch1970,hansen1974,thorne1980}.  Attempting a similar strategy here, one might begin by noting that, in Newtonian physics, the field is given by
\begin{align}\label{PhiNewt}
\Phi = \frac{1}{2} \mE_{ij} x^i x^j - \frac{M}{r} - \frac{3}{2} \frac{Q_{ij}n^i n^j}{r^3}.
\end{align}
where we introduce $n^i=x^i/r$.  Given that $-(1+g_{00})/2$ plays the role of a Newtonian potential, one might propose to write
\begin{align}\label{GRfail}
-\frac{1+g_{00}}{2} = \frac{1}{2} \mE_{ij} x^i x^j + \dots + \frac{-3}{2} \frac{Q_{ij}n^i n^j}{r^3} + O\left( \frac{1}{r^4} \right),
\end{align}
where the dots ($\dots$) allow for intervening powers of $r$ (i.e. relativistic corrections) between the $r^2$ tidal field and the $r^{-3}$ response.  In this proposal one solves a relativistic stellar structure model in the presence of a tidal field, and infers the induced quadrupole moment from the coefficient of the $1/r^3$ term of the time-time component of the metric.  Unfortunately, this strategy is just as ambiguous as the first: What coordinate $r$ is to be used?  For example, defining \cite{fang-lovelace2005}
\begin{align}\label{FL}
r' = r\left[1+N \left(\frac{M}{r}\right)^5 \right]
\end{align} 
for some number $N$, the induced moment changes by
\begin{align}
Q_{ij}' = Q_{ij} + \tfrac{2}{3} N M^5 \mE_{ij},
\end{align}
making the deformabilities change by
\begin{align}\label{ambig}
\lambda' = \lambda - \tfrac{2}{3} N M^5, \qquad k' = k - N \mathcal{C}^5, \qquad \Lambda' = \Lambda - \tfrac{2}{3} N.
\end{align}
The full ambiguity is discussed in Sec.~\ref{sec:coordinate} below.  

\section{Relativistic Framework: Detailed Discussion}\label{sec:relativistic}

To lay a foundation for a more careful discussion, I now briefly review of the formal setup for considering a small body in general relativity.  More detailed and precise treatments may be found in (e.g.) \cite{thorne-hartle1985,gralla-wald2008,pound2010}.  We consider a body of typical radius $R$ immersed in an external universe of typical scale of variation $\mathcal{R} \gg R$ and define $\epsilon =R/\mathcal{R} \ll 1$.  Since both scales are important, the problem can be treated in the langauge of matched asymptotic expansions.  For $r \ll \mathcal{R}$, we have the near expansion ($\epsilon \rightarrow 0$ fixing $R$),
\begin{align}
g_{\mu \nu} = g^{\rm near}_{\mu \nu} + h_{\mu \nu}^{\rm near} + O(\epsilon^2).
\end{align}
We will see that $g^{\rm near}_{\mu \nu}$ is an asymptotically flat metric representing the gravitational field of the compact object in isolation, while $h_{\mu \nu}^{\rm near}$ is a small correction due to the tidal field.  Formally, these terms are of order $\epsilon^0$ and $\epsilon^1$, respectively, in the near expansion.  Both are important in this analysis.

For $r \gg R$, we have the far expansion ($\epsilon \rightarrow 0$ fixing $\mathcal{R}$),
\begin{align}
g_{\mu \nu} = g^{\rm far}_{\mu \nu} + h_{\mu \nu}^{\rm far} + O(\epsilon^2).
\end{align}
Here $g^{\rm far}_{\mu \nu}$ is a vacuum spacetime representing the external universe in which the body is embedded, while $h_{\mu \nu}^{\rm near}$ is a small correction due to the body field.  Formally, these terms are of order $\epsilon^0$ and $\epsilon^1$, respectively, in the far expansion.  Only the background metric $g^{\rm far}_{\mu \nu}$ is important in this analysis.  The correction $h_{\mu \nu}^{\rm far}$ accounts for self-force effects.

We assign the body a geodesic $\gamma$ of the far metric $g_{\mu \nu}^{\rm far}$.  The metric can be expanded in local inertial coordinates $(t,x^i)$ where $t=0$ is the geodesic and $g_{\mu \nu}= \eta_{\mu \nu} + O(r^2)$ with  $r=\sqrt{\delta_{ij}x^i x^j}$.  We use the THZ coordinates \cite{thorne-hartle1985,zhang1986,detweiler2005}, in which
\begin{subequations}\label{THZ}
\begin{align}
g^{\rm far}_{00} & = -1 - \mE_{ij} x^i x^j + O(r^3) \\
g^{\rm far}_{i0} & = \frac{2}{3} \epsilon_{ijk} x^j \mB^k_{\ l} x^l + O(r^3) \\
g^{\rm far}_{ij} & = \delta_{ij}( 1 - \mE_{kl}x^k x^l ) + O(r^3).
\end{align}
\end{subequations}
Here $\mE_{ij}(t)$ and $\mB_{ij}(t)$ are symmetric trace-free tensor fields, whose indices are raised and lowered with $\delta_{ij}$.  Computing the Riemann tensor of this metric, we find
\begin{equation}
\mathcal{E}_{ij} = R^{\rm ext}_{0i0j}|_{\gamma} , \qquad \mathcal{B}_{ij}  = -\frac{1}{2} \epsilon^{kl}_{\ \ i}R^{\rm ext}_{0jkl}|_{\gamma},
\end{equation}
illustrating that $\mE_{ij}$ and $\mB_{ij}$ capture the ten components of the Riemann/Weyl tensor of the vacuum external universe.

Near a single time $t_0$ (i.e. $|t-t_0| \ll \mathcal{R}$), we may match in the region of overlap $R \ll r \ll \mathcal{R}$ and thereby constrain the near-zone metric.  One finds that the near-zone background and perturbation are both stationary.  The background is furthermore asymptotically flat,
\begin{align}\label{gnear}
g^{\rm near}_{\mu \nu} = \eta_{\mu \nu} + O\left(\frac{1}{r}\right),
\end{align}
while for the perturbation we must have
\begin{subequations}\label{hnear}
\begin{align}
h^{\rm near}_{00} & = - \mE_{ij} x^i x^j + O(r) \label{hnear00} \\
h^{\rm near}_{i0} & = \frac{2}{3} \epsilon_{ijk} x^j \mB^k_{\ l} x^l + O(r) \\
h^{\rm near}_{ij} & = - \mE_{kl}x^k x^l \delta_{ij} + O(r).
\end{align}
\end{subequations}
Notice the consistency of \eqref{gnear} and \eqref{hnear00} with Eq.~\eqref{GRfail} above.  As in the Newtonian case, we assume that the body is initially spherical (i.e., $g^{\rm near}_{\mu \nu}$ is static and spherically symmetric).  We will also consider the case where the gravitomagnetic tidal field vanishes, $\mB_{ij}=0$.  We discuss generalizations in Sec.~\ref{sec:differences} below.

For a given stellar model, our task is to solve the coupled matter and Einstein equations in the near expansion with boundary conditions \eqref{hnear} (here taking $\mB_{ij}=0$).  Outside the star where the background metric is exactly Schwarzschild, the matter degrees of freedom are irrelevant and $h^{\rm near}_{\mu \nu}$ solves vacuum linearized Einstein equation in the Schwarzschild spacetime.  Only the $\ell=2$, even-parity modes of the star will couple to the boundary conditions \eqref{hnear}, and hence only $\ell=2$, even-parity modes will be present in $h_{\mu \nu}^{\rm near}$.  The general solution outside the star takes the form
\begin{align}\label{CC}
h^{\rm near}_{\mu \nu} = h^{\rm H}_{\mu \nu} + C h^{\infty}_{\mu \nu},
\end{align}
where $C$ is a constant.  The first solution $h^{\rm H}_{\mu \nu}$ is regular on the horizon but not at infinity, while the second solution $h^{\infty}_{\mu \nu}$ is regular at infinity but not on the horizon.  These conditions define the two solutions invariantly up to overall normalization.  We choose the normalization to be adapted to the tidal field boundary condition, i.e.,
\begin{subequations}\label{norm}
\begin{align}
h^{\rm H}_{00} & = -\mE_{ij} x^i x^j + O(r) \label{hinf00} \\
h^{\infty}_{00} & = 3 \mE_{ij} n^i n^j \frac{M^5}{r^3} + O\left( \frac{1}{r^4} \right).\label{hH00}
\end{align}
\end{subequations}
This choice completes the definition of the constant $C$, in particular making it dimensionless.  Full expressions for  $h^{\rm H}_{\mu \nu}$  and $h^{\infty}_{\mu \nu}$ in a convenient coordinate system are given in the appendix.

\subsection{Coordinate-dependent approach}\label{sec:coordinate}

To connect to the quasi-Newtonian approach \eqref{GRfail}, we note from Eqs.~\eqref{thing1} and \eqref{thing2} that in the areal coordinate system and RWZ gauge we have simply
\begin{align}
h^{\rm H}_{00} & = -\mE_{ij} n^i n^j  \left( r-2M \right)^2.
\end{align}
In the areal coordinate system and RWZ gauge, the horizon-regular solution $h^{\rm H}_{00}$ is an infinite series in $1/r$, while the infinity-regular solution $h^{\rm H}_{00}$ happens to terminate before the $r^{-3}$ term is reached.  Thus the only contribution to the $r^{-3}$ term is from $h^{\infty}_{00}$, which is normalized so that the tidal deformability is just
\begin{align}\label{ARWZ}
\textrm{Areal coordinates, RWZ Gauge:  \ \ \ } \Lambda = C.
\end{align}
In other coordinate systems, the tidal deformability identified this way would be different.  In a new coordinate $r'$ related to the areal coordinate $r$ by
\begin{align}\label{rp}
r = r'\left[1+\alpha_0 \frac{M}{r'} + \alpha_1 \left( \frac{M}{r'} \right)^2 + \alpha_2 \left( \frac{M}{r'} \right)^3 + \dots \right],
\end{align}
the horizon-regular solution $h^{H}_{00}$ now makes a contribution to the $r^{-3}$ term, and the new deformability is
\begin{align}\label{lambdaprime}
\Lambda' = C - \tfrac{2}{3} \left( \alpha_2 \alpha_3  - 2 \alpha_4 + \alpha_1 \alpha_4  + \alpha_5 \right).
\end{align}
However, $\Lambda$ turns out to be invariant under the most common changes of coordinates (isotropic coordinates, harmonic coordinates, light-cone gauge), perhaps explaining why discrepancies have not yet arisen in practice.  Note that the transformation \eqref{lambdaprime} takes a different form if the initial coordinates are not areal and RWZ.  There is no simple, general transformation law for $\Lambda$ as defined in this way.  The ambiguity in the naive tidal deformability is essentially as big as the diffeomorphism freedom itself.

\subsection{Solution-dependent approach}\label{sec:solution}

We noted above that Eq.~\eqref{CC}, together with conditions listed below, invariatly defines a constant $C$ for each stellar model.  Since $C$ is invariant and reduces to the tidal deformability in the Newtonian limit, why not take this number to be the relativistic tidal deformability?  But such a definition would be rather unusual, because it makes the \textit{neutron star} tidal deformability depend in an essential way on \textit{black hole} physics.  In effect, a physicist attempting to determine the induced quadrupole from the metric near a tidally-perturbed neutron star is asked to analytically continue the field towards a fictitious black hole horizon $r_H$ and carefully examine the blow-up of the field to discern the coefficient $C$.  A definition based on Eq.~\eqref{CC}---a convenient split for black hole physics---unduly privileges the black hole in a definition that is supposed to apply to all compact objects. Indeed, it leads directly to the conclusion that the black hole tidal deformability vanishes.

This type of definition could similarly be made with reference to a different family of compact objects paramterized by the mass $M$.  To pick a definite example, let us suppose that APR-stars \cite{APR} are to be afforded this privileged status.   We can then invariantly define solutions $\hat{h}^{\mathcal{O}}_{\mu \nu}$ and $\hat{h}^{\infty}_{\mu \nu}$ together with a constant $\hat{C}$ by
\begin{align}\label{CC2}
h^{\rm near}_{\mu \nu} = \hat{h}^{\mathcal{O}}_{\mu \nu} + \hat{C} \hat{h}^{\infty}_{\mu \nu},
\end{align}
together with (i) both solutions are $\ell=2$ and even parity; (ii) $\hat{h}^{\mathcal{O}}_{\mu \nu}$ is regular at the origin when continued into an APR-star; (iii) $\hat{h}^{\infty}$ is regular at infinity; (iv) the normalization choice analogous to \eqref{norm} is made,
\begin{subequations}\label{norm}
\begin{align}
\hat{h}^{\mathcal{O}}_{00} & = -\mE_{ij} x^i x^j + O(r) \label{hinf00} \\
\hat{h}^{\infty}_{00} & = 3 \mE_{ij} n^i n^j \frac{M^5}{r^3} + O\left( \frac{1}{r^4} \right).\label{hH00}
\end{align}
\end{subequations}
 This defines a dimensionless constant $\hat{C}$ for any stellar model, which reduces to the tidal deformability in the Newtonian limit.  Is this, then, the relativistic tidal deformability?  Of course not.  A definition based on Eq.~\eqref{CC2}---a convenient split for APR-star physics---unduly privileges the APR-star in a definition that is supposed to apply to all compact objects. Indeed, it leads directly to the conclusion that the APR-star tidal deformability vanishes. 

In the solution-dependent approach, the ambiguity is one choice of a preferred family of compact objects.

\subsection{Size of the ambiguity}

Since tidal deformability is well-defined in the Newtonian limit, the ambiguity must become less important as the star becomes less compact.  We can therefore estimate the size of the ambiguity in terms of the stellar compactness $\mathcal{C}=M/R$ by considering the  limit $\mathcal{C} \rightarrow 0$.   Since $k$ is the natural dimensionless parameter in the Newtonian setting, we can anticipate from simple analysis [e.g. Eq.~\eqref{ambig}] that the relative size of the ambiguity will be of order $\mathcal{C}^5$.

More formally, we can take the Newtonian limit by letting $\mathcal{C} \rightarrow 0$ at fixed $M$, $\mE_{ij}$, and $\bar{r} \equiv r/R$.  This means that the radius of the star becomes large, $R \to \infty$, but the observer stays a proportionate distance away.  It is well known that for Newtonian stellar models the deformability $\lambda$ scales like $R^{5}$, so we assume that the Love number $k \sim \Lambda R^{-5}$ has a non-zero, finite limit.  A valid Newtonian limit must use a coordinate $r$ such that the Newtonian potential \eqref{PhiNewt} is properly reproduced.  The areal-coordinate, RWZ gauge solution \eqref{CC} gives
\begin{align}\label{nearNewt}
\lim_{\mathcal{C} \to 0} h^{\rm near}_{00} = - \mE_{ij}  n^i n^j \bar{r}^2\left(1 + \frac{2k}{\bar{r}^{5}} \right) .
\end{align}
Since there are no intervening powers of $\bar{r}$ between $\bar{r}^2$ and $\bar{r}^{-3}$, we see that these coordinates properly reproduce the Newtonian limit.  To understand the coordinate freedom we can write Eq.~\eqref{rp} as
\begin{align}\label{rp2}
\bar{r} = \bar{r}'\left[1+\alpha_1 \frac{\mathcal{C}}{\bar{r}'} + \alpha_2 \left( \frac{\mathcal{C}}{\bar{r}'} \right)^2 + \dots \right],
\end{align}
which (using \eqref{lambdaprime} with $k=(3/2) \Lambda \mathcal{C}^5$) changes the Love number by 
\begin{align}\label{loveydovey}
k' = k+ \left( \alpha_2 \alpha_3  - 2 \alpha_4 + \alpha_1 \alpha_4  + \alpha_5 \right)  \mathcal{C}^{5}.
\end{align}
If the coefficients $\alpha_i$ are simply numbers, then we recover the expected scaling $\mathcal{C}^5$ of the ambiguity.  However, in principle the coefficients can depend on the perturbation parameter $\mathcal{C}$; this is the usual gauge freedom of perturbation theory.  The requirement the Newtonian limit \eqref{nearNewt} be preserved gives only the mild restriction that $\lim_{\mathcal{C}\to 0}\alpha_n \mathcal{C}^n = 0$.  For example, one could choose $\alpha_5=\mathcal{C}^{-4}$ without messing up the Newtonian limit.  In this case the Love number changes by order $\mathcal{C}$ instead of $\mathcal{C}^5$, a much larger effect.  However, this seems rather contrived as, in effect, the Love number changes its definition for each radius of star.  If we add the stipulation to that the coordinates are to be defined from the geometry outside the star,\footnote{The areal coordinate $r$ is defined intrinsically by the geometry outside the star, as is the mass $M$.  Thus any coordinate $r'=r f(r/M)$ for smooth $f$ can be considered defined by the asymptotic geometry.} then the coefficients $\alpha_2$ are fixed in the Newtonian limit, making the ambiguity again of the expected order $\mathcal{C}^5$. 

Having established that the coefficients $\alpha_i$ should be fixed in the Newtonian limit, one can take these as order-$1$ numbers for estimating the actual size of the ambiguity.  The statement is simplest in terms of $\Lambda$, which just shifts by products of the $\alpha_i$ [Eq.~\eqref{lambdaprime}].  These products might range from $.1$ to $10$ in practice, so we can estimate the size of the ambiguity as:
\begin{align}
\Lambda \rightarrow \Lambda + \textrm{(number of order .1 to 10)}
\end{align}
For neutron stars, $\Lambda$ typically ranges from $\sim 50$ to $\sim 2000$ (e.g. \cite{hinderer-lackey-lang-read2010}), so this ambiguity is potentially important only for the least deformable models.  For black holes, the ambiguity is essentially order unity.  The fact that $\Lambda_{\rm BH}=0$ using the standard definition can be interpreted as stating that the tidal deformability $\Lambda_{\rm BH}$ of a black hole is of order unity, as expected on dimensional grounds.  More theoretical work is required before gravitational-wave observations can constrain properties of black holes or other very compact objects via their tidal deformability.

\section{Tidal Deformability Differences}\label{sec:differences}

We have discussed two ways to fix the ambiguity, by adopting a preferred choice of coordinates or a preferred family of compact objects.  I now propose a third way to view the ambiguity, which is really just a variation on the second theme.  The key observation is that, while the metric of a tidally-perturbed star is (by definition) not asymptotically flat, the \textit{difference} between the metric perturbations of two equal-mass objects subject to the same tidal field is asymptotically flat.  To see this we return to \eqref{CC}, which defines constants $C_1$ and $C_2$ for compact obects $1$ and $2$, both of mass $M$, whose tidal perturbations are encoded in metric perturbations ${}_1 h_{\mu \nu}^{\rm near}$ and ${}_2 h_{\mu \nu}^{\rm near}$.  Subtracting the perturbations gives
\begin{align}
\delta h_{\mu \nu}^{\rm near} = {}_2 h_{\mu \nu}^{\rm near} - {}_1 h_{\mu \nu}^{\rm near} = \left( C_2 - C_1 \right) h^{\infty}_{\mu \nu},
\end{align}
where we remind that the solution $h^{\infty}_{\mu \nu}$ is asymptotically flat,
\begin{align}
h^{\infty}_{00} = 3 \mE_{ij} n^i n^j \frac{M^5}{r^3} + O\left( \frac{1}{r^4} \right).
\end{align}
The quadrupole moment of the perturbation difference $\delta h_{\mu \nu}^{\rm near}$ is thus unambiguous, and the ``tidal deformability difference'' is just
\begin{align}
\delta \Lambda = C_2 - C_1.
\end{align}

Having established the basic properties using a particular coordinate system, gauge, and set of basis solutions to the linearized Einstein equation, we can now state the definition invariantly as follows:

\begin{itemize}
\item[] \begin{mdframed} Consider the metric of a static, spherically symmetric body perturbed by an $\ell=2$, even-parity tidal field.  Consider a second such metric for a different body of the same mass $M$ in the same tidal field $\mE_{ij}$.  In a neighborhood of infinity (i.e. outside the stars), these metrics take the form 
\begin{align}
{}_1 g_{\mu \nu} & = g^{\rm Schw}_{\mu \nu} + {}_1 h_{\mu \nu} \\
{}_2 g_{\mu \nu} & = g^{\rm Schw}_{\mu \nu} + {}_2 h_{\mu \nu},
\end{align} 
where $g^{\rm Schw}_{\mu \nu}$ is the Schwarzschild metric.  Since the background spacetimes are diffeomorphic, we may identify them unambiguously and consider the perturbations ${}_1 h_{\mu \nu}$ to live on the same manifold.  Then we may construct a new metric by
\begin{align}
{}_{21} g_{\mu \nu} = g^{\rm Schw}_{\mu \nu} + {}_{21} h_{\mu \nu}, \qquad {}_{21} h_{\mu \nu} \equiv {}_{2} h_{\mu \nu} - {}_{1} h_{\mu \nu}.
\end{align}
The analysis above shows that there exists a gauge where the perturbation ${}_{21} h_{\mu \nu}$ falls off like $1/r^3$.  That is, ${}_{21} g_{\mu \nu}$ is an asymptotically flat, vacuum solution to Einstein's equation (through first order in perturbation theory), and hence has a well-defined set of multipole moments \cite{geroch1970,hansen1974,thorne1980,gursel1983}.  In particular, its mass quadrupole moment ${}_{21} Q_{ij}$ is well defined.  We say that this moment is the \textit{induced quadrupole difference} associated with compact objects $1$ and $2$.  The tidal deformability difference parameters are defined relative to ${}_{21} Q_{ij}$ in the usual way [i.e., Eq.~\eqref{lambda}, \eqref{k}, or \eqref{Lambda}].
\end{mdframed}
\end{itemize}
This definition has some nice features.  First, its statement does not require the use of coordinates or a preferred solution to the linearized Einstein equation, making it immediately clear that it is free of the ambiguities discussed above.  Second, it generalizes immediately to tidal deformability parameters associated with arbitrary multipoles \cite{binnington-poisson2009,damour-nagar2009}, since the general structure of perturbations of Schwarzschild implies that the difference perturbation will still be asymptotically flat.  It likely also applies to slowly rotating bodies (of the same mass and spin), or in general to two bodies with the same multipole moments. To confirm, one must verify (using Einstein's equation) that the difference metric is indeed asymptotically flat.  This should follow directly from the statement that solutions are uniquely characterized, in a neighborhood of infinity, by their multipole moments \cite{beig-simon1980}.

Of course, this approach does not define the tidal deformability of a single body; instead it suggests that only the difference of deformabilities is meaningful.  One could define the tidal deformability of single bodies by adopting an additional axiom, such as
\begin{itemize}
\item \textit{(Axiom.)}  The tidal deformability of a black hole vanishes.
\end{itemize}  
This axiom, combined with the above definition of tidal deformability differences, reproduces the standard definition of relativistic tidal deformability \cite{hinderer2008,binnington-poisson2009,damour-nagar2009}.

\section{Previous Work}\label{sec:comparison}

Relativistic tidal deformabilities were first computed by Hinderer \cite{hinderer2008}, who adopted the approach of Sec.~\ref{sec:coordinate}, using areal coordinates and RWZ gauge, as in Eq.~\eqref{ARWZ}.  Later, independent work of Damour and Nagar \cite{damour-nagar2009} and Binnington and Poisson \cite{binnington-poisson2009} generalized to arbitrary (linear) tidal fields.  These works give clear, unambiguous (and equivalent) definitions of the tidal deformability parameters, with a slight difference in attitude toward the results.  In particular, Binnington and Poisson \cite{binnington-poisson2009} regard the vanishing of the black hole Love numbers as a ``correct interpretation of [the] results'', while Damour and Nagar \cite{damour-nagar2009} regard it as a ``consistency check on [the] formal definition''.  
In both works, the definition is clearly presented in terms of the split \eqref{CC}, i.e. in the spirit of what I have called the solution-dependent approach in Sec.~\ref{sec:solution}.  The issue was later revisited in an effective field theory framework by Kol and Smolkin \cite{kol-smolkin2012}, who motivate the split by an analytic continuation of the spacetime dimension.  Generalizations to spinning bodies (beyond the scope of this work) have subsequently been given in Refs.~\cite{landry-poisson2015,pani-gualtieri-maselli-ferrari2015,pani-gualtieri-ferrari2015,landry2017}.

The first extensive discussion of the ambiguity in separating a tidal response from the applied tidal field was given by Fang and Lovelace \cite{fang-lovelace2005}, who gave the coordinate-transformation argument reviewed in Eqs.~\eqref{FL}-\eqref{ambig} above. 
Binnington and Poisson \cite{binnington-poisson2009} note that their definition is free of this ambiguity since their background coordinates are geometrically fixed.  Their framework allows only infinitesimal coordinate transformations, which change the deformability only at second order in perturbation theory and can be neglected.  This is a resolution of the ambiguity in that a specific, unambiguous definition is adopted; however, one may still regard the notion of deformability as ambiguous in that other reasonable definitions are possible.  The size of the ambiguity was estimated in footnote 12 of Ref.~\cite{pani-gualtieri-ferrari2015}, whose conclusions are consistent with mine.  The purpose of this paper is not to wordsmith definitions or argue for any particular one, but to suggest a new way of thinking about deformability---in terms of differences---that may inform the conversation about waveform modeling.

\section{The Waveform}\label{sec:waveform}

I have argued that the notion of relativistic tidal deformability is inherently ambiguous.  The ambiguity can be fixed by adopting a preferred choice of coordinates or a preferred family of compact objects, but my personal choice is to focus only on the unambiguous \textit{difference} in tidal deformabilities.  Ultimately, however, the best definition is the one that best facilitates the process of linking of observable quantities to compact object microphysics.  I now speculate on how the notion tidal deformability difference may be of use in determining equation-of-state effects on gravitational waveforms.

Consider a gravitational waveform generated from a compact binary system in the PN regime.  That is, the members of the binary can be strong-field objects like black holes, but the orbital separation must be large, such that the typical velocity is much less than the speed of light.  As a model, we can use a self-consistent PN waveform (up to some fixed order like 3PN) together with the replacement 
\begin{align}
Q_{ij}^{\rm PN} = -  M^5 \Lambda^{\rm GW} \mE^{\rm PN}_{ij},
\end{align} 
where $Q_{ij}^{\rm PN}$ and $\mE^{\rm PN}_{ij}$ are the quadrupole moment and tidal tensor defined in the PN expansion.  (Of course, one needs such a term for each of the two bodies.)  Formally, these terms are of higher (5PN) order and should not be included.  By including them we provide a \textit{definition} of the tidal deformability $\Lambda^{\rm GW}$.  This parameter can be measured for a given waveform by fitting to the PN model, as done recently by the LIGO collaboration \cite{LIGONS}.

I have argued that this observable tidal deformability $\Lambda^{\rm GW}$ is not guaranteed to agree with the tidal deformability $\Lambda^{\rm standard}$ as normally defined and calculated \cite{hinderer2008,binnington-poisson2009,damour-nagar2009}, which makes use of a conventional choice with no obvious counterpart in the waveform model.  Instead, I propose to reinterpret $\Lambda^{\rm standard}$ as the (convention-independent) \textit{difference} between the tidal deformability of the object and that of a black hole of the same mass.  Although I cannot prove it, it seems clear that the PN waveform definition $\Lambda^{\rm GW}$ corresponds to an allowed convention in this framework.  In this case, the relationship to the standard deformability is simply
\begin{align}\label{Lambdaoffset}
\Lambda^{\rm GW} = \Lambda^{\rm standard} + \Lambda^{\rm GW}_{\rm BH},
\end{align}
where $\Lambda^{\rm GW}_{\rm BH}$ is the PN-defined tidal deformability of a black hole.  One can then measure $\Lambda^{\rm GW}_{\rm BH}$, once and for all, by doing a high-resultion simulation of a binary black hole system in the PN (well-separated) regime and fitting the waveform to the PN model.  The $\Lambda^{\rm GW}$ for (say) a neutron star can then be computed from the standard tidal deformability together with Eq.~\eqref{Lambdaoffset}, and the neutron star waveform, complete with tidal effects, can be generated with some confidence.

This strategy need not be limited to PN waveforms.  The basic observation is that, when tidal deformabilities are included in a waveform model in a reasonable way, one can expect the ambiguity in relating back to the equation of state to be a single number that can be measured by fitting the model to a binary black hole waveform.  More traditional semi-analytic methods  (e.g., \cite{bernuzzi-dietrich-damour2015,hinderer-etal2016,dietrich-bernuzzi-tichy2017, lackey-bernuzzi-galley-vdb2017}) rely on numerical simulations of neutron star mergers for calibration.  The method proposed here calibrates with vacuum (black hole) simulations, which are computationally cheaper and hence more practical to perform in the challenging early-inspiral regime.


An approach more similar to ours is the ``tidal splicing'' method of Ref.~\cite{barkett-etal2016}.  The idea is to produce a neutron star waveform by adding tidal terms to a numerically-generated binary black hole waveform.  In this case, the black hole tidal deformability is already included in the waveform, so the deformability parameters in the spliced terms represent the tidal deformability \textit{difference} between a neutron star and a black hole.  I have argued that the standard definition of tidal deformability corresponds to precisely this difference.  Thus, the analysis of this paper suggests that the $\lambda$ of Ref.~\cite{barkett-etal2016} is equal to the tidal deformability as usually defined \cite{binnington-poisson2009,damour-nagar2009}, lending support to the tidal splicing method.

The method I suggest also offers some advantages over tidal splicing.  In the splicing approach, generating a waveform with a given set of parameters requires interpolating between a bank of numerically-generated waveforms before adding in the tidal terms.  In the approach suggested here, by contrast, one need calculate a single number $\Lambda^{\rm GW}_{\rm BH}$ from a single high-accuracy simulation, after which all waveforms can be efficiently generated from the analytical model.  Of course, one would want to perform several simulations to ensure that $\Lambda^{\rm GW}_{\rm BH}$ does not depend on the binary parameters (mass ratio, eccentricity, etc.).  This can be considered a test of whether the given waveform model is suitable for applying the method.\footnote{If $\Lambda^{\rm GW}_{\rm BH}$ is found to depend on system parameters, then it cannot be equal to an allowed choice of tidal deformability parameter in the present framework, since here the tidal deformability is defined as an intrinsic property of the compact object, without reference to the specifics of its tidal environment.}  But once the value (and constancy) of $\Lambda^{\rm GW}_{\rm BH}$ is established, waveform generation becomes fully analytic, offering a computationally efficient method for including neutron star microphysics in binary inspiral waveforms.



\section*{Acknowledgements}

I wish to thank Eanna Flanagan and Eric Poisson for helpful correspondence.  This work was supported by NSF grant 1506027 to the University of Arizona.

\appendix
\section{Perturbations of Schwarzschild}

The static vacuum perturbations of Schwarzschild are treated in many references.  We find the expressions in Ref.~\cite{field-hesthaven-lau2010} particularly useful for writing down the multipolar solutions.  Here we present $\ell=2$, even-parity solution in terms of the Schwarzschild isotropic coordinate $\rho$ and in ``RWZ gauge''.  We introduce two linearly independent solutions $h^{\rm H}$ and $h^{\rm \infty}$, which are regular at the horizon and at infinity, respectively.  These are normalized according to Eq.~\eqref{norm}.  The complete expression is
\begin{subequations}
\begin{align}
h^{\rm H/\infty}_{00} & =\mE_{ij} n^i n^j \mathcal{F}^{\rm H/\infty}(\rho)  \label{thing1} \\
h^{\rm H/\infty}_{i0} & = 0 \\ 
h^{\rm H/\infty}_{ij} & = \mE_{kl}n^k n^l \left[ \delta_{ij}\mathcal{G}^{\rm H/\infty}(\rho) +   n_i  n_j   \mathcal{H}^{\rm H/\infty}(\rho)\right],
\end{align}
\end{subequations}
where
\begin{align}
\mathcal{F}^{\rm H}(\rho)  & = -\frac{(M-2 \rho )^4}{16 \rho ^2} \label{thing2} \\
\mathcal{G}^{\rm H}(\rho)  & = -\frac{(M+2 \rho )^4 \left(M^4+8 M^3 \rho -8 M^2 \rho ^2+32 M \rho ^3+16 \rho ^4\right)}{256 \rho ^6} \\
\mathcal{H}^{\rm H}(\rho)  & = \frac{M (M+2 \rho )^4 \left(M^2+4 \rho ^2\right)}{32 \rho ^5}
\end{align}
and
\begin{align}
\mathcal{F}^{\infty}(\rho)  & = -\frac{15}{256 \rho ^2 (M+2 \rho )^4} \Bigg[ 8 \left(3 M^7 \rho -44 M^5 \rho ^3-176 M^3 \rho ^5+192 M \rho ^7\right) \nonumber \\ & + 3 \left(M^2-4 \rho ^2\right)^4 \log \left(\frac{(M-2 \rho )^2}{(M+2 \rho )^2}\right)\Bigg] \\
\mathcal{G}^{\infty}(\rho)  & = -\frac{15 (M+2 \rho )^2}{4096 \rho ^6} \Bigg[8 M \rho  \left(3 M^4+36 M^3 \rho +88 M^2 \rho ^2+144 M \rho ^3+48 \rho ^4\right) \nonumber \\ & + 3 \left(M^4+8 M^3 \rho -8 M^2 \rho ^2+32 M \rho ^3+16 \rho ^4\right) (M+2 \rho )^2 \log \left(\frac{(M-2 \rho )^2}{(M+2 \rho )^2}\right)\Bigg]\\
\mathcal{H}^{\infty}(\rho)  & = -\frac{3 (M+2 \rho )^2}{4096 \rho ^6 (M-2 \rho )^2} \Bigg[3 M^8+144 M^7 \rho -48 M^6 \rho ^2-1856 M^5 \rho ^3+288 M^4 \rho ^4 \nonumber \\ & -7424 M^3 \rho ^5-768 M^2 \rho ^6+9216 M \rho ^7+768 \rho ^8 \nonumber \\ & +15 \left(M^2-4 \rho ^2\right)^4 \log \left(\frac{(M-2 \rho )^2}{(M+2 \rho )^2}\right)\Bigg].
\end{align}
The isotropic coordinate is convenient because all of the horizon-regular functions are terminating polynomials. The areal coordinate is related by $r=\rho(1+M/(2\rho))^2$.  \\

\bibliographystyle{utphys}
\bibliography{love}

\end{document}